\documentclass{article}
\usepackage{multirow}
\usepackage{graphicx}
\usepackage{biblatex}
\usepackage[a4paper,
            bindingoffset=0.2in,
            left=1in,
            right=1in,
            top=1in,
            bottom=1in,
            footskip=.25in]{geometry}
\usepackage[utf8]{inputenc}
\usepackage{color,soul}
\usepackage{caption}
\usepackage{ragged2e}
\usepackage{subcaption}
\usepackage{arydshln}

\newcommand\blfootnote[1]{%
  \begingroup
  \renewcommand\thefootnote{}\footnote{#1}%
  \addtocounter{footnote}{-1}%
  \endgroup
}
\bibliography{references.bib}
\title{Domain Shift Analysis in Chest Radiographs Classification in a Veterans Healthcare Administration  Population  }

\author{Mayanka Chandrashekar$^1$$^*$, Ian Goethert$^2$, Md Inzamam Ul Haque$^1$, \\ Benjamin McMahon$^4$, 
Sayera Dhaubhadel$^4$, Kathryn Knight$^4$,  Joseph Erdos$^{3,5}$, \\ Donna Reagan$^3$, Caroline Taylor$^3$, Peter Kuzmak$^3$, \\ John Michael Gaziano $^{3,5}$,
Eileen McAllister$^2$, Lauren Costa$^3$,\\ Yuk-Lam Ho$^3$, Kelly Cho$^3$,
Suzanne Tamang $^6$, Samah Fodeh-Jarad $^5$, \\ Olga S. Ovchinnikova$^1$, Amy C. Justice$^{3,5}$,  Jacob Hinkle$^1$, Ioana Danciu$^{1,7}$}

\begin{document}

\date{}
% \twocolumn
% [
%   \begin{@twocolumnfalse}
    \maketitle
\parbox{\textwidth}{% 
    \centering
     $^1$ Advanced Computing for Health Sciences, Computational Sciences and Engineering Division,\\ Oak Ridge National Laboratory \\
     $^2$ IT Services Division, Oak Ridge National Laboratory \\
     $^3$ Veteran Affair's Association \\
     $^4$ Los Alamos National Laboratory \\
     $^5$ Yale University \\
     $^6$ Stanford University \\
     $^7$ Department of Biomedical Informatics, Vanderbilt University Medical Center \\
     $^*$ Corresponding Author Email: chandrashekm@ornl.gov
     }

\begin{abstract}

\textbf{Objectives}: This study aims to assess the impact of domain shift on chest X-ray classification accuracy and to analyze the influence of ground truth label quality and demographic factors such as age group, sex, and study year.

\textbf{Materials and Methods}: We used a DenseNet121 model pretrained MIMIC-CXR dataset for deep learning-based multilabel classification using ground truth labels from radiology reports extracted using the CheXpert and CheXbert Labeler. We compared the performance of the 14 chest X-ray labels on the MIMIC-CXR and Veterans Healthcare Administration chest X-ray dataset (VA-CXR). The VA-CXR dataset comprises over 259k chest X-ray images spanning between the years 2010 and 2022. 

\textbf{Results}:
The validation of ground truth and the assessment of multi-label classification performance across various NLP extraction tools revealed that the VA-CXR dataset exhibited lower disagreement rates than the MIMIC-CXR datasets. Additionally, there were notable differences in AUC scores between models utilizing CheXpert and CheXbert. When evaluating multi-label classification performance across different datasets, minimal domain shift was observed in unseen datasets, except for the label "Enlarged Cardiomediastinum." The study year's subgroup analyses exhibited the most significant variations in multi-label classification model performance. These findings underscore the importance of considering domain shift in chest X-ray classification tasks, particularly concerning study years.

\textbf{Conclusion}:
Our study reveals the significant impact of domain shift and demographic factors on chest X-ray classification, emphasizing the need for improved transfer learning and equitable model development. Addressing these challenges is crucial for advancing medical imaging and enhancing patient care.

\end{abstract}

% \end{@twocolumnfalse}
% ]
{\noindent\justifying{\blfootnote{ Notice: Office of Science of the U.S. Department of Energy. This manuscript has been authored by UT-Battelle, LLC, under contract DE-AC05-00OR22725 with the US Department of Energy (DOE). The US government retains and the publisher, by accepting the article for publication, acknowledges that the US government retains a nonexclusive, paid-up, irrevocable, worldwide license to publish or reproduce the published form of this manuscript, or allow others to do so, for US government purposes. DOE will provide public access to these results of federally sponsored research in accordance with the DOE Public Access Plan (http://energy.gov/downloads/doe-public-access-plan). }}}

\clearpage

\section{Introduction}

Chest radiography is the first-line imaging test for respiratory disease and some forms of cardiac disease. Chest X-ray abnormalities detection has been automated by recent advanced artificial intelligence techniques. Accurate chest X-ray classification has played an important role in many biomedical applications to accelerate the diagnosis and treatment of conditions like pneumonia, heart failure, rib trauma, pulmonary fibrosis, etc. Though there has been an increase in the integration of machine learning models to enhance diagnostic promise, the efficacy of these models hinges on the availability and quality of training data. 

Recent years have seen a wave of artificial intelligence (AI) and machine learning models for clinical applications from large research medical centers or based on limited de-identified datasets. The introduction of transfer learning has resulted in AI models being accessible to researchers with fewer computational resources and less data, allowing them to leverage existing knowledge. Transfer learning is the ability to apply a model trained on a dataset to another dataset of interest.  A major consideration for transfer learning is domain shift, the dissimilarity between data distributions from the source used for training and the population to which the models are applied. This divergence between publicly available datasets and private, institution-specific datasets can introduce substantial bias and hinder the generalization of machine learning models to real-world scenarios.

In this study, we quantify the domain shift between the public domain (MIMIC-CXR) and a private dataset (VA-CXR). The efficacy of the transfer learning approach for chest X-ray classification and the subsequent impact on classification accuracy when dealing with domain shift form the main focus of this study.  The domain shift is traditionally viewed only based on the model and its performance, ignoring the multi-fold causes leading to the shift. We comprehensively address the effects of domain shift in this three-stage study: 1) Compare the performance between the source domain and target domain accuracy on the chest X-ray classification. 2) Quantify the quality of the ground truth extracted from the radiology reports, as supervised learning models are heavily dependent on the quality of the labels.  3) Analyze the relationship between demographic factors and classification accuracy, as domain mismatches often originate in demographic mismatches. 

By addressing the critical interplay of domain shift and demographic factors, this study not only provides insights into the technical challenges of adapting machine learning models to private datasets but also underscores the broader implications of these challenges in the context of healthcare. Our research contributes to developing more accurate, robust, and generalizable chest X-ray classification models by creating a systematic approach for using an existing model and understanding the nuances of a model's performance before applying it to a new population.

\section{Methods}
\begin{figure*}
    \centering
    \includegraphics[width=\linewidth]{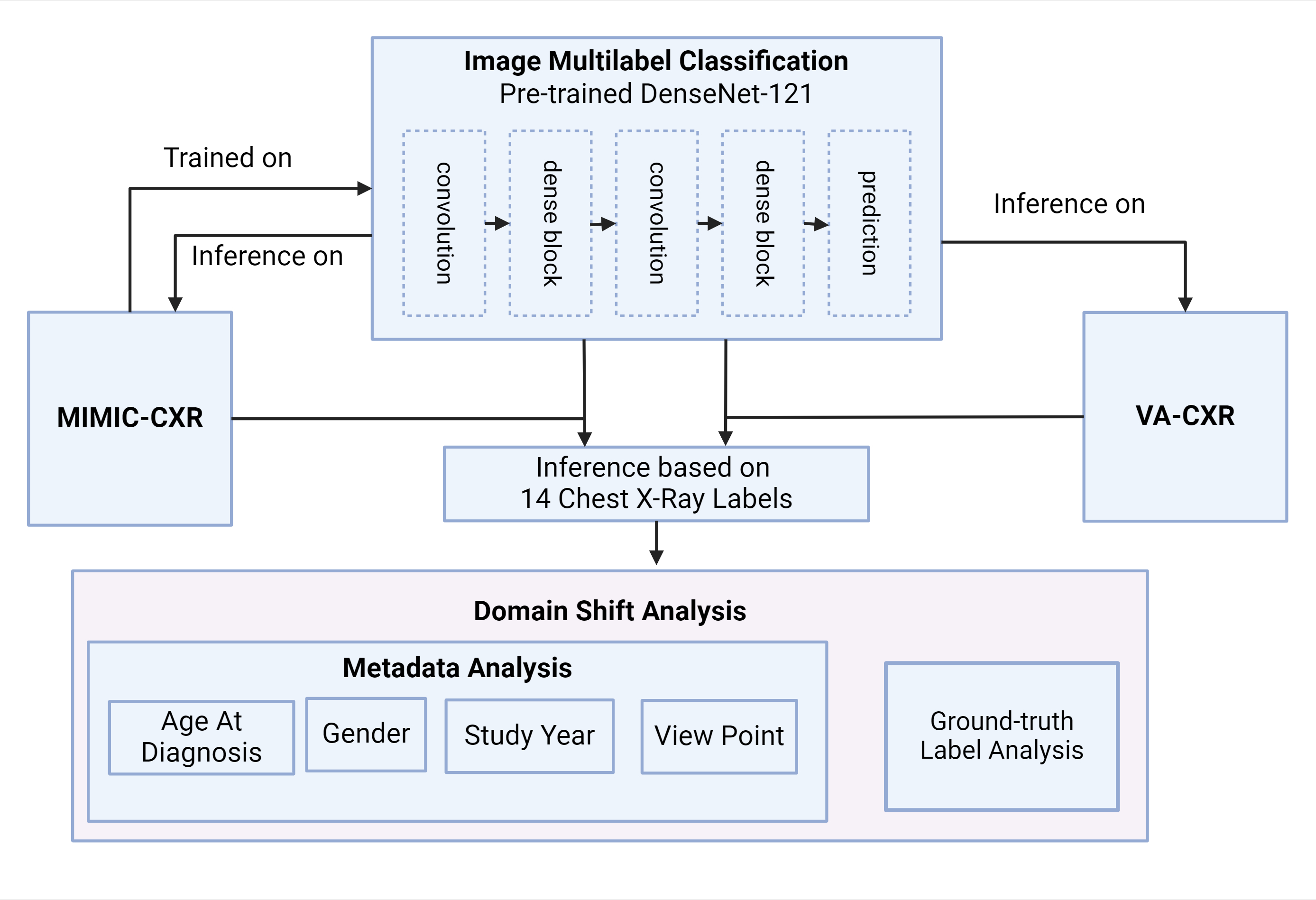}
    \caption{Domain Shift Analysis based on Image Classification Model}
    \label{fig:arch}
\end{figure*}

\subsection{Dataset}
\noindent \textbf{MIMIC-CXR} is a publicly available dataset of 377,110 chest X-rays associated with 227,827 imaging studies from 65,379 patients. MIMIC-CXR is collected from the inpatient setting of Beth Israel Deaconess, a Boston hospital. (refer Table~\ref{tab:datasets}) We also use the test-split of this dataset (Test-split MIMIC-CXR) as the hold-out set analysis, which consists of 5159 studies with 293 patients \cite{johnson2019mimicjpg}. 

\noindent \textbf{VA-CXR} is a private dataset of 259,361 chest X-rays associated with 91,020 imaging studies from 35,771 patients. VA-CXR is collected in the outpatient setting of the Boston Veterans Healthcare Administration station (refer to Table~\ref{tab:datasets}). The ground truth labels were extracted from the VA's corporate data warehouse (CDW) and joined with images using patient information from DICOM headers as published in Knight et al. \cite{knight2024vision}.

\begin{table*}[h!]
\centering 
\renewcommand{\arraystretch}{1.2}
\begin{tabular}{|c||cc||cc|}
\hline
& \multicolumn{2}{|c|}{\textbf{Source Dataset}}      & \multicolumn{2}{|c|}{\textbf{Target Dataset}}  \\ \hline
\textbf{Name}         & \multicolumn{2}{|c|}{MIMIC-CXR}              & \multicolumn{2}{|c|}{VA-CXR} \\ \hline
\textbf{Availability} & \multicolumn{2}{|c|}{Public}             & \multicolumn{2}{|c|}{Private}  \\ \hline

\textbf{Clinical Setting} &  \multicolumn{2}{|c|}{Inpatient} &  \multicolumn{2}{|c|}{Outpatient}\\ \hline
\textbf{Time}         & \multicolumn{2}{|c|}{2011 to 2016}   & \multicolumn{2}{|c|}{2010 to 2022}     \\ \hline
\textbf{\# Images}    & \multicolumn{2}{|c|}{377,110} & \multicolumn{2}{|c|}{259,361}    \\ \hline
\textbf{\# Studies}   & \multicolumn{2}{|c|}{227,827} & \multicolumn{2}{|c|}{91,020}   \\ \hline 
\textbf{\# Patients} & \multicolumn{2}{|c|}{65,379} & \multicolumn{2}{|c|}{35,771} \\ \hline

\multicolumn{5}{|c|}{\textbf{\# Studies by Age (Percentage \%)}} \\ \hline
(0-50{]}    &  51607 &  (22.65)  & 8589  & (9.43)  \\ \hline
(50-60{]}   & 41661  &  (18.29)  & 12258 & (13.46) \\ \hline
(60-70{]}   &  51168 &  (22.46)  & 30736 & (33.76) \\ \hline
(70-80{]}   & 43060  &  (18.9)   & 23360 & (25.66) \\ \hline
(80-90{]}   &  30854 &  (13.54)  & 13229 & (14.53) \\ \hline
(90-100{]}  &  9477  &  (4.16)   & 2825  & (3.10)  \\ \hline

\multicolumn{5}{|c|}{\textbf{\#Patients by Sex (Percentage \%)}} \\ \hline
\textbf{Male} &  34,127  & (52.2) & 33,381 & (93.31) \\ \hline
\textbf{Female} & 31,006  & (47.4) & 2,390 & (6.68) \\ \hline
\textbf{NaN}    & 246 & (3.76) &  - & - \\ \hline
\end{tabular}

\caption{Source and Target Dataset}
\label{tab:datasets}
\end{table*}

\subsection{Ground truth Label Extraction}

We used the CheXbert labelers \cite{smit2020chexbert} to expertly assign labels to 14 specific labels (Atelectasis, Cardiomegaly, Consolidation, Edema, Enlarged Cardiomediastinum, Fracture, Lung Lesion, Lung Opacity, No Finding, Pleural Effusion, Pleural Other, Pneumonia, Pneumothorax, Support Devices) associated with different chest conditions from radiology reports. CheXbert is a BERT-based approach that automates the detection of these observations, effectively streamlining the process of annotating medical images and reports. The NLP label extraction outputs scores for four classes: positive, negative, blank, and uncertain, associated with each of the 14 labels. As the class names indicate, for example, for \textit{Pneumonia}, positive class: radiology report indicates that the patient has \textit{Pneumonia}; negative class: radiology report indicates that the patient doesn't have  \textit{Pneumonia}; uncertain class: radiology report mentions  \textit{Pneumonia} but NLP tool unable to determine if it is positive or negative; blank class: radiology report doesn't mention  \textit{Pneumonia}.

\subsection{Label Validation}
Because ground truth assignment ultimately determines the accuracy of the imaging classifiers, we developed an image evaluation procedure in two steps. First, we evaluated the agreement between the NLP label extraction tool, CheXbert, and its precursor, CheXpert \cite{irvin2019chexpert}, a rule-based tool. We focused on positive class agreements for our evaluation, using only positive classes for classification, and have combined uncertain/negative class agreements. The disagreement between NLP label extraction tools indicates ambiguity/ less confidence on the labels, ultimately creating an unreliable ground truth. The agreement is measured for both MIMIC-CXR and VA-CXR datasets; Of note that neither of the datasets was used to train the NLP extraction tools.

\subsubsection{Relation to Diagnoses Codes}
To validate the ground extracted from chexpert-labeler, we analyzed the relationship of specific ground truth labels to ICD codes in the patient's electronic health record (EHR) extracted from the VA's Corporate Data Warehouse (CDW).The assignment of ICD-9 and ICD-10 diagnosis codes associated with each condition was exploratory and not extensively optimized. 

Starting concepts were retrieved from the Chexpert-labeler github repository , where phrases they used to search notes can be found at \textit{https://github.com/stanfordmlgroup}.  These phrases, along with our own expertise, were used to identify diagnosis codes and cross-reference radiology reports with diagnoses.For example:   Pneumonia was identified by chexpert-labeler as indicated by \textit{pneumonia}, \textit{infection}, \textit{infected process}, and \textit{infectious}; 
Edema was indicated by terms \textit{edema}, \textit{heart failure}, \textit{chf}, \textit{vascular congestion}, \textit{pulmonary congestion}, \textit{indistinctness}, and \textit{vascular prominence}; Fracture was indicated solely by the word \textit{fracture}; and Pneumothorax was identified by either \textit{pneumothorax} or \textit{pneumothoraces}.

This method enabled us to validate the ground truth labels by correlating them with the relevant ICD codes in the patients' EHRs, ensuring accurate cross-referencing of radiology reports with diagnoses.

For instance, the ICD-9 codes we used to indicate a pneumonia diagnosis in the outpatient diagnosis tables ranged from 480 to 486 and included 487.0. These codes encompass viral, bacterial, and other types of pneumonia, as well as pneumonia caused by unspecified pathogens. A similar approach was applied to ICD-10 codes for pneumonia and other conditions. The specific ICD codes used for each condition are detailed in Appendix Table~\ref{tab:ground_icd}.

\subsection{Multi-label Image Classification}

Using the 14 labels extracted from the corresponding radiology reports with CheXbert, we created a multi-label image classification model from X-ray images. We used a pre-trained DenseNet model\cite{huang2017densely} as the core framework, removed the top classification layer, and integrated a custom classification layer for multi-label output. 
Previous work shows the effectiveness of different resolutions of DenseNet-121 based multi-label classification chest X-ray model on MIMIC-CXR dataset  \cite{haque2023effect}. This work uses the MIMIC-CXR trained DenseNet-121 Model on Chest X-ray pre-processed into 256x256 JPG images \cite{johnson2019mimicjpg, johnson2019mimic}.

\subsection{Metrics}
We evaluated our models using the Area Under the Curve (AUC). The AUC score was calculated separately for each of the 14 labels, indicating the separability measure for a given chest X-ray label. 
We also analyzed the difference in AUC scores between MIMIC-CXR and VA-CXR, and the prevalence for each label was calculated as the number of studies with positive results for a given label divided by the total number of labels, indicating the label's presence in the given cohort.

\subsection{Domain Shift Analysis}
We compared our source and target datasets along different dimensions: 1) Demographic Details: Age At Time of Imaging Study, Sex; 2) Imaging Study Details: Study Year, View  Point (Lateral view (Lat), erect anteroposterior (AP), posteroanterior (PA)) ; 3) Ground truth Labels: 14 labels. All the factors were analyzed against the accuracy of multi-label image classification. 
The impact can be estimated by comparing the prevalence of the above-listed factors across source and target domains with the performance of the multi-label classification. 
% \subsection{Sub-group Analysis}
The subgroup analysis is performed on unseen datasets: Test Split MIMIC-CXR and VA-CXR.

\noindent \textbf{Study Year Analysis}: We obtain the study year from VA-CXR studies date reported in DICOM headers.  The MIMIC-CXR dataset is unidentified; study years are replaced by anchor years. The comparison of performance across study years between the datasets isn't possible, so we don't present a study year analysis on the MIMIC-CXR dataset.

\noindent \textbf{Performance Analysis based on Sex}: We obtain the sex of the patients in VA-CXR from the VA's corporate data warehouse (CDW) and the patients in MIMIC-CXR from the metadata. Sex is only classified into binary classes of male and female. We analyze the label-wise performance for each sex across both datasets.

\noindent \textbf{View Position Analysis}: We obtain the view position of the images from the DICOM metadata. We use four main viewpoints for analysis: Lateral view (Lat), erect anteroposterior (AP), posteroanterior (PA))  and Left Lateral view (LL). We compare the label-wise performance across both datasets

\noindent \textbf{Age Group Analysis}:
We define the age of the patient as the age when the imaging study was performed. For VA-CXR, this age was calculated from the date of birth obtained from VA CDW and the date of imaging study from DICOM metadata. For MIMIC-CXR, we calculated age using the anchored date of birth and anchored imaging study date. The anchored dates were amended in a manner that the difference in years is constant, so we were able to approximate the age of the patient. The label-wise performance for both datasets was compared across six age groups: (0-50], (50-60], (60,70], (70,80], (80,90], and (90,100]. 

\section{Results}

\subsection{Ground truth Label Validation}
Table~\ref{tab:Groundtruth} presents a comprehensive analysis of agreement and disagreement rates between Chexpert and Chexbert on the MIMIC-CXR and VA-CXR datasets across the 14 labels. The results shed light on the performance and consistency of Chexpert and Chexbert and the uncertainty of the ground truth labels. Notably, in MIMIC-CXR, Atelectasis was identified in 19.5\% of cases with a disagreement rate of 10.1\%, whereas in VA-CXR, the identification rate was lower at 9.8\% with a disagreement rate of only 0.7\%. This discrepancy in positive identification and disagreement rates is further exemplified in conditions like Cardiomegaly, where MIMIC-CXR reported positive identification in 15.7\% of cases with a significant disagreement rate of 39.1\%, contrasting with VA-CXR's 9.3\%  positive identification rate and 13.7\% disagreement rate.

\begin{table}[h]
\renewcommand{\arraystretch}{1.5}
\footnotesize
\begin{tabular}{|l|l|l|l|l|l|l|l|l|l|l|l|l|}
\hline
\multirow{3}{*}{\textbf{Labels}} & \multicolumn{6}{l|}{\textbf{MIMIC-CXR}}   & \multicolumn{6}{l|}{\textbf{VA-CXR}}   \\ \cline{2-13}
                                      & \multicolumn{2}{p{2cm}|}{\textbf{Positive Agreement}}  & \multicolumn{2}{p{2cm}|}{\textbf{U/N* Agreement}} & \multicolumn{2}{p{2cm}|}{\textbf{Disagree- ment}}   & \multicolumn{2}{p{2cm}|}{\textbf{Positive Agreement}}  & \multicolumn{2}{p{2cm}|}{\textbf{U/N* Agreement}}  & \multicolumn{2}{p{2cm}|}{\textbf{Disagree- ment}}  \\ \cline{2-13}
                                      & \textbf{Count}     & \textbf{\%} & \textbf{Count}               & \textbf{\%} & \textbf{Count}         & \textbf{\%} & \textbf{Count}    & \textbf{\%} & \textbf{Count}               & \textbf{\%} & \textbf{Count}         & \textbf{\%} \\ \hline
\textbf{Atelectasis}                  & 44422              & 19.5        & 8557                         & 3.76        & 23028                  & 10.1        & 8902              & 9.8         & 4454                         & 4.9         & 621                    & 0.7         \\ \hline
\textbf{Cardiomegaly}                 & 35733              & 15.7        & 19798                        & 8.69        & 88995                  & 39.1        & 8437              & 9.3         & 29427                        & 32.3        & 12459                  & 13.7        \\ \hline
\textbf{Consolidation}                & 9785               & 4.3         & 10868                        & 4.77        & 54675                  & 24.0        & 1306              & 1.4         & 24019                        & 26.4        & 4863                   & 5.3         \\ \hline
\textbf{Edema}                        & 25559              & 11.2        & 32134                        & 14.10       & 39799                  & 17.5        & 2104              & 2.3         & 35534                        & 39.0        & 2864                   & 3.1         \\ \hline
\textbf{ECM}                          & 4612               & 2.0         & 13365                        & 5.87        & 95973                  & 42.1        & 2487              & 2.7         & 27615                        & 30.3        & 14264                  & 15.7        \\ \hline
\textbf{Fracture}                     & 3920               & 1.7         & 797                          & 0.35        & 10941                  & 4.8         & 4049              & 4.4         & 677                          & 0.7         & 588                    & 0.6         \\ \hline
\textbf{Lung Lesion}                  & 5769               & 2.5         & 973                          & 0.43        & 8877                   & 3.9         & 5408              & 5.9         & 1497                         & 1.6         & 2295                   & 2.5         \\ \hline
\textbf{Lung Opacity}                 & 44982              & 19.7        & 2065                         & 0.91        & 30895                  & 13.6        & 16965             & 18.6        & 9696                         & 10.7        & 19734                  & 21.7        \\ \hline
\textbf{No Finding}                   & 18951              & 8.3         & 0                            & 0.00        & 56921                  & 25.0        & 12654             & 13.9        & 0                            & 0.0         & 3770                   & 4.1         \\ \hline
\textbf{PE}             & 52084              & 22.9        & 28513                        & 12.52       & 100762                 & 44.2        & 11804             & 13.0        & 59301                        & 65.2        & 4729                   & 5.2         \\ \hline
\textbf{PO}                & 1960               & 0.9         & 645                          & 0.28        & 3761                   & 1.7         & 4192              & 4.6         & 544                          & 0.6         & 959                    & 1.1         \\ \hline
\textbf{Pneumonia}                    & 8054               & 3.5         & 36599                        & 16.06       & 40233                  & 17.7        & 2351              & 2.6         & 3913                         & 4.3         & 3083                   & 3.4         \\ \hline
\textbf{PT}                 & 8461               & 3.7         & 41010                        & 18.00       & 100543                 & 44.1        & 3344              & 3.7         & 34191                        & 37.6        & 2780                   & 3.1         \\ \hline
\textbf{SD}              & 63526              & 27.9        & 827                          & 0.36        & 28196                  & 12.4        & 13294             & 14.6        & 233                          & 0.3         & 8221                   & 9.0         \\ \hline
\end{tabular}
\caption{Groundtruth Analysis: Agreement and Disagreement Rates between ChexPert and ChexBert across 14 classification labels \\ {\footnotesize *Uncertain/ Negative; PE-Pleural Effusion; PT- Pneumothorax ; SD - Support Devices; PO- Pleural Other;*ECM - Enlarged Cardiomediastinum}}

\label{tab:Groundtruth}  
\end{table}

\begin{table}[h]
 \renewcommand{\arraystretch}{1.1}
\centering
\begin{tabular}{|c|c|c|c|c|c|c|} \hline
                                    & \multicolumn{3}{|c|}{\textbf{ChexPert}}               & \multicolumn{3}{c|}{\textbf{ChexBert}}               \\ \hline
\textbf{Finding}                    & \textbf{AUC} & \textbf{Prevalence} & \textbf{Count} & \textbf{AUC} & \textbf{Prevalence} & \textbf{Count} \\ \hline
\textbf{Atelectasis}                & 0.801        & 0.101               & 21329          & 0.802        & 0.100               & 21169          \\ \hline
\textbf{Cardiomegaly}               & 0.753        & 0.149               & 31461          & 0.862        & 0.100               & 21151          \\ \hline
\textbf{Consolidation}              & 0.746        & 0.025               & 5265           & 0.811        & 0.016               & 3308           \\ \hline
\textbf{Edema}                      & 0.794        & 0.035               & 7453           & 0.854        & 0.026               & 5528           \\ \hline
\textbf{ECM*} & 0.519        & 0.161               & 33970          & 0.600        & 0.035               & 7466           \\ \hline
\textbf{Fracture}                   & 0.619        & 0.047               & 9910           & 0.622        & 0.046               & 9623           \\ \hline
\textbf{Lung Lesion}                & 0.647        & 0.066               & 13975          & 0.652        & 0.069               & 14609          \\ \hline
\textbf{Lung Opacity}               & 0.716        & 0.260               & 54872          & 0.736        & 0.199               & 41965          \\ \hline
\textbf{No Finding}                 & 0.784        & 0.165               & 34771          & 0.790        & 0.156               & 32883          \\ \hline
\textbf{Pleural Effusion}           & 0.920        & 0.136               & 28770          & 0.927        & 0.141               & 29689          \\ \hline
\textbf{Pleural Other}              & 0.746        & 0.046               & 9806           & 0.739        & 0.055               & 11703          \\ \hline
\textbf{Pneumonia}                  & 0.683        & 0.055               & 11533          & 0.755        & 0.030               & 6306           \\ \hline
\textbf{Pneumothorax}               & 0.868        & 0.043               & 8973           & 0.898        & 0.037               & 7845           \\ \hline
\textbf{Support Devices}            & 0.793        & 0.211               & 44596          & 0.804        & 0.161               & 33935         \\ \hline
\end{tabular}
\caption{ Multi-Label Image Classification using MIMIC-CXR(plus Chexpert) as Source, Results shown in table shows the performance of VA-CXR on two NLP Ground Truth Extraction for VA-CXR  \\ {\footnotesize *ECM - Enlarged Cardiomediastinum}}
\label{tab:GroudtruthPerformance}
\end{table}

\subsection{Comparison with Diagnosis Codes}

While the X-ray classification accuracy was evaluated against the label extracted using NLP of the appropriate radiology report, it is possible also to compare directly against diagnosis codes in the clinical record.  However, interpreting this comparison is challenging due to two main issues: first, diagnoses are based on a broader range of information beyond just the X-ray, and second, patients may have multiple conditions, making it difficult to determine the specific reason each X-ray was ordered. Despite these challenges, we have conducted two useful comparisons of X-ray findings against diagnosis codes for nine categories within our dataset.

\begin{figure*}
    \centering
    \includegraphics[width=.7\linewidth]{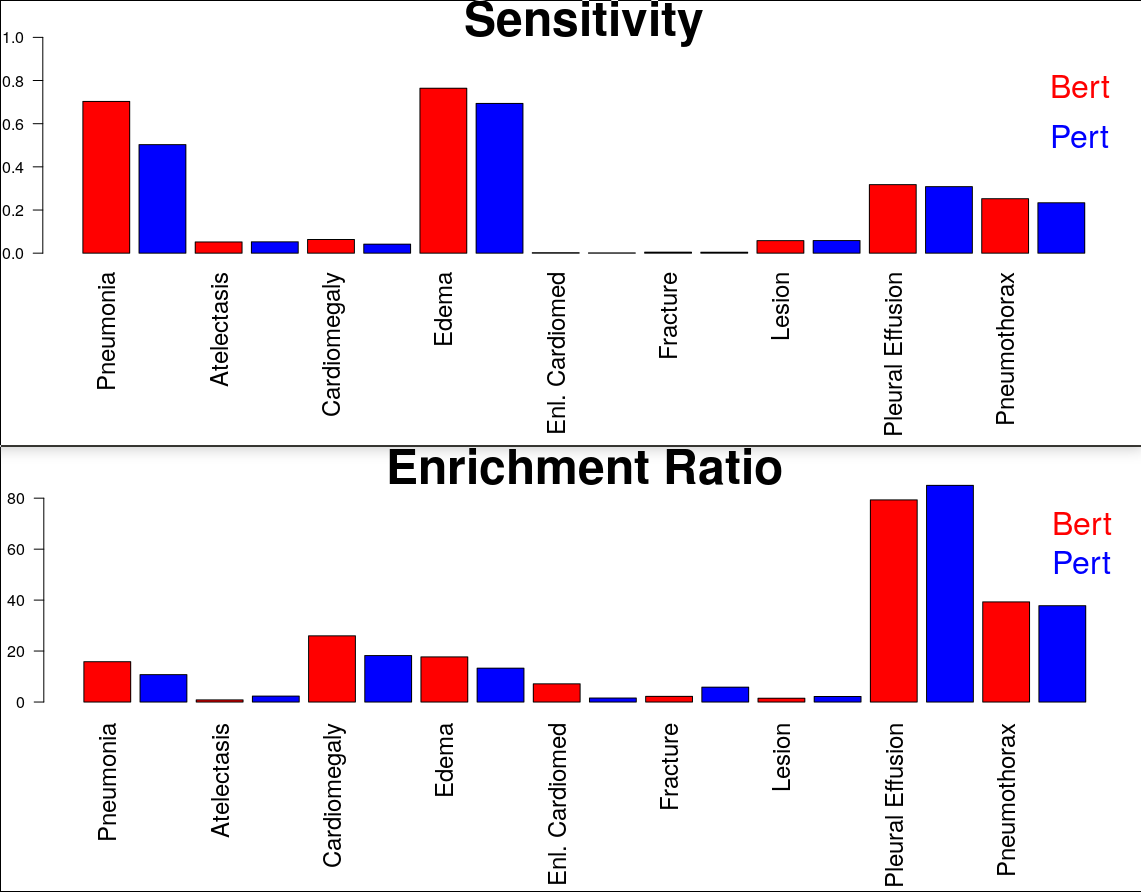}
    \caption{\centering Characterization of the concurrence of positive findings on chest X-rays with associated diagnoses for our VA data set.  The top panel shows the sensitivity with which a related diagnosis is given within a week before or after a positive x-ray finding, which we labeled 'Sensitivity'.  The bottom panel shows the factor by which the ratio of positive to negative X-ray findings increases when a diagnosis code is present.  Patients with a diagnosis code not within one week of the assessed X-ray are excluded from the calculation for both plots. 
    {\footnotesize  {Bert- Chex-Bert Model ; Pert- Chex-Pert Model }}}
    \label{fig:sensDx.png}
\end{figure*}

The top of Figure \ref{fig:sensDx.png} shows the fraction where a diagnosis related to the label was recorded within a week (either before or after) of the X-ray, compared to the total number of patients, including those who never had the diagnosis. Patients who had the diagnosis but not within the one-week window were excluded from this calculation. This plot, labeled 'Sensitivity,' reflects the proportion of cases where a diagnosis occurs within a week of noting the condition in the X-ray.  
We observe a positive finding on an X-ray, as extracted from the radiology report, which is mostly associated with a specific diagnosis of pneumonia and edema and frequently associated with pleural effusion and pneumothorax.  We also see that the ChexBert model is more frequently associated with a corresponding diagnosis than the ChexPert model.

The bottom of Figure \ref{fig:sensDx.png} compares the factor by which the enrichment ratio of positive finding in the radiology report increases with a concurrent (within one week) diagnosis.  Pleural effusion, for example, is seen 10 times more frequently in the radiology report when a concurrent diagnosis code is noted.  It is also true, however, that 2/3 of the pleural effusion diagnoses are not accompanied by a positive finding in a radiology report, as shown in the top bar chart, and also that 90\% of the time a negative finding is made, no diagnosis is found.  We see that the conditions with the highest enrichment ratios differ from those with the highest sensitivities, but the trend that the ChexBert model generally outperforms the ChexPert model by a few to 20\% remains.

Several factors were observed to contribute to the quantitative comparison of labels extracted from the radiology reports to the observed diagnoses, including the extent to which diagnosis codes could be found that correspond well to the radiology report finding, whether the X-ray is a screening or confirmatory test, the prevalence of the condition, and the chronic vs. acute nature of the condition.  We provide this figure as a survey across these factors and present the complete set of counts in Appendix Table~\ref{tab:ground_count}.  Specific ICD 9 and 10 codes used for each condition are provided in the Appendix Table~\ref{tab:ground_icd}.

\subsection{Multi-label Image Classification on VA-CXR across NLP tools}

Table~\ref{tab:GroudtruthPerformance} compares ground truth extraction techniques, specifically focusing on the label-wise performance metrics on the VA-CXR dataset. Two techniques, ChexPert and ChexBert, are evaluated based on their Area Under the Curve (AUC), prevalence, and count across 14 labels. Across the findings, both techniques generally demonstrate similar AUC values, indicating comparable performance in distinguishing positive cases. For instance, in identifying Atelectasis, both ChexPert and ChexBert exhibit AUC values around 0.80, with similar prevalence and count numbers. Notably, in the case of Cardiomegaly, ChexBert shows a notably higher AUC of 0.862 compared to ChexPert's 0.753, suggesting potentially superior performance in this specific finding. However, the prevalence and count metrics vary across the findings and between the two techniques. For example, ChexPert tends to have higher prevalence and count values for several findings like ECM (Enlarged Cardiomediastinum), while ChexBert shows higher values for others such as Pleural Effusion.

\subsection{Multi-label Image Classification Performance across multiple datasets}

Table~\ref{tab:datasetAUC} shows the comparison of MIMIC-CXR (Source dataset), Test Split MIMIC-CXR (Hold-out Source Dataset), and VA-CXR based on AUC on the 14 labels.  The Hold-out Source dataset is the test split of MIMIC-CXR dataset,  DenseNet-121 Model. The Test Split of MIMIC-CXR gives us a fair comparison to VA-CXR, the unseen target dataset. This can be observed based on the AUC drop from the overall MIMIC-CXR to test-split. In Table~\ref{tab:datasetAUC}, the difference in AUC between Hold-out and Target indicates the performance variation between VA-CXR and Test Split of MIMIC. The Negative value of the difference in AUC indicates that the VA-CXR performs better than the Test Split MIMIC-CXR, and the positive value indicates that the Test Split performs better. The Enlarged Cardiomediastinum (ECM) label has the highest difference in AUC, indicating a huge performance drop in VA-CXR. This could directly impact the lack of a large number of image studies in ECM in the source dataset compared to the target. 

\begin{table}[h]
\renewcommand{\arraystretch}{1.2}
\centering
 \begin{tabular}{|p{3cm}|r|r||r|r|r|r|r|} \hline 
\multirow{3}{*}{ \textbf {Finding } }         &  \multicolumn{2}{|c||}{\textbf{MIMIC-CXR}}&  \multicolumn{2}{|p{3cm}|}{\centering\textbf{Test-Split MIMIC-CXR}}&  \multicolumn{2}{|c|}{\textbf{VA-CXR}} & \multicolumn{1}{p{2cm}|}{\multirow{3}{=}{\centering\textbf{Difference in AUC  between Hold-out and Target}}}\\ \cline{2-7}   
         & \multicolumn{2}{|c||}{\textbf{Source Dataset}} & \multicolumn{2}{|p{3cm}|}{\centering\textbf{Hold-out Source Dataset}} & \multicolumn{2}{c|}{\textbf{Target Dataset}} & \\ \cline{2-7}
        &  \textbf {AUC}       & \textbf {Count} &  \textbf {AUC}                  &  \textbf{Count} &  \textbf {AUC}    &  \textbf {Count} &                   \\ \hline
 \textbf {Atelectasis }      & 0.808                               & 45808                           & 0.762                                          & 1034                            & 0.801                            & 21329                           & -0.039            \\ \hline
 \textbf {Cardiomegaly }     & 0.816                               & 44845                           & 0.793                                          & 1258                            & 0.753                            & 31461                           & 0.04              \\ \hline
 \textbf {Consolidation }    & 0.821                               & 10778                           & 0.762                                          & 326                             & 0.746                            & 5265                            & 0.016             \\ \hline
 \textbf {Edema }            & 0.889                               & 27018                           & 0.832                                          & 959                             & 0.794                            & 7453                            & 0.038             \\ \hline
 \textbf {Enlarged Cardio-mediastinum }  & 0.739                   & 7179                            & 0.727                                          & 200                             & 0.519                            & 33970                           & 0.208             \\ \hline
 \textbf {Fracture }         & 0.667                               & 4390                            & 0.714                                          & 167                             & 0.619                            & 9910                            & 0.095             \\ \hline
 \textbf {Lung Lesion }      & 0.738                               & 6284                            & 0.722                                          & 202                             & 0.647                            & 13975                           & 0.075             \\ \hline  \textbf {Lung Opacity }     & 0.749                               & 51525                           & 0.705                                          & 1561                            & 0.716                            & 54872                           & -0.011            \\ \hline
 \textbf {No Finding }       & 0.853                               & 75455                           & 0.809                                          & 983                             & 0.784                            & 34771                           & 0.025             \\ \hline
 \textbf {Pleural Effusion } & 0.919                               & 54300                           & 0.894                                          & 1542                            & 0.920                            & 28770                           & -0.026            \\ \hline
 \textbf {Pleural Other }    & 0.806                               & 2011                            & 0.806                                          & 119                             & 0.746                            & 9806                            & 0.06              \\ \hline
 \textbf {Pneumonia }        & 0.714                               & 16556                           & 0.713                                          & 539                             & 0.683                            & 11533                           & 0.03              \\ \hline
 \textbf {Pneumothorax }     & 0.856                               & 10358                           & 0.813                                          & 144                             & 0.868                            & 8973                            & -0.055            \\ \hline
 \textbf {Support Devices }  & 0.898                               & 66558                           & 0.876                                          & 1457                            & 0.793                            & 44596                           & 0.083       \\ \hline      
\end{tabular}
\caption{ Comparison of Labelwise Accuracy between Source, Hold-out Source and Target Datasets\\
{\footnotesize *Enlarged Cardio-mediastinum}} 
\label{tab:datasetAUC}
\end{table}
\clearpage

\subsection{Study Year-wise Performance on VA-CXR }
\begin{figure*}[h]
    \centering
    \includegraphics[width=\linewidth]{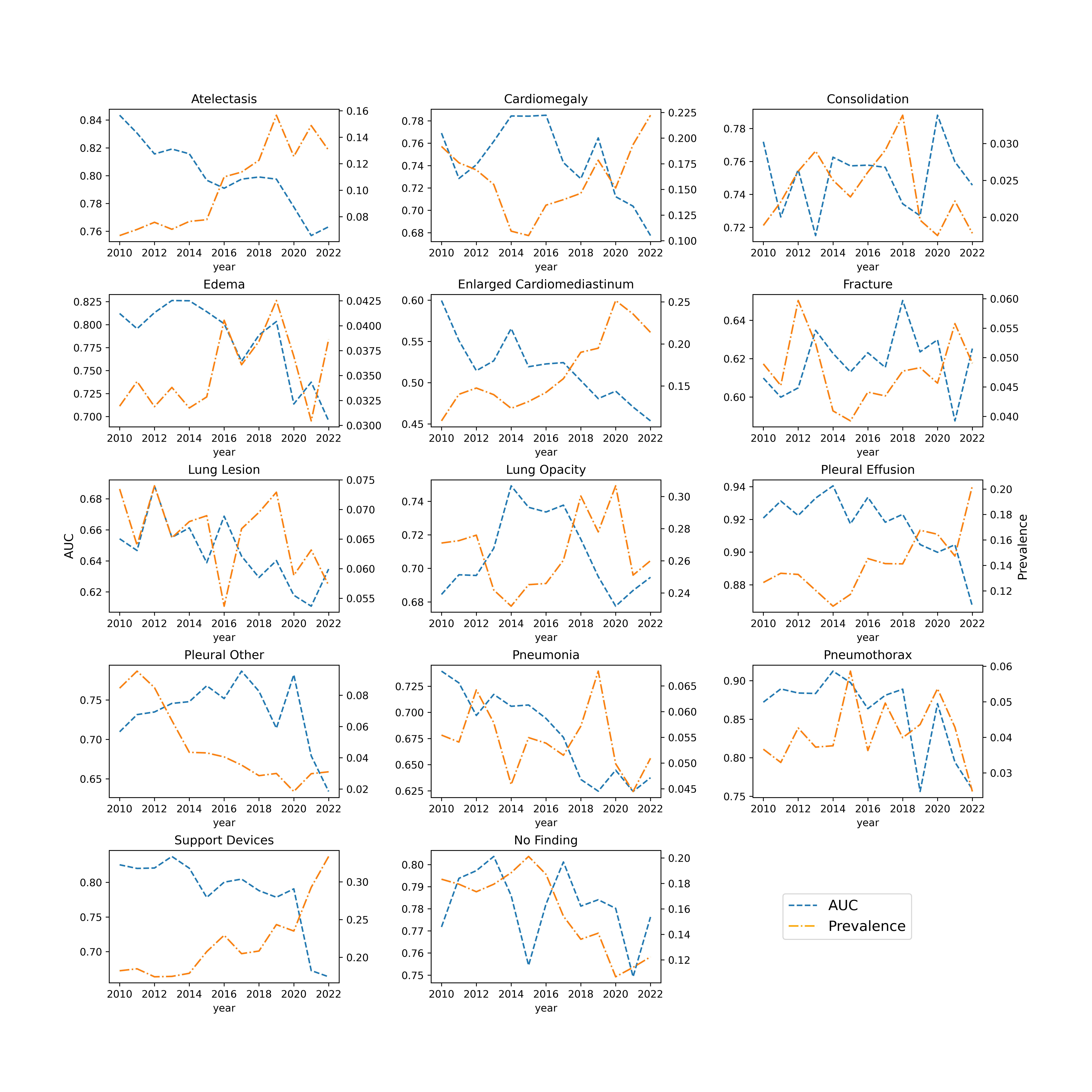}
    \caption{ Accuracy and Prevalence of Labels Study Year Wise \\ {\footnotesize *Blue line indicates the AUC across years. Orange Line indicates Prevalence across years}}
    
    \label{fig:studyYear}
\end{figure*}

Table~\ref{fig:studyYear} shows the label-wise distribution of the study years. The AUC systematically drops as for study year \textit{2020 to 2022} for all labels except \textit{Consolidation} and \textit{Pleural Other}, which peaks at \textit{2020}. The prevalence increases over the years in VA-CXR for \textit{Atelectasis}, \textit{Enlarged Cardiomediastinum}, and \textit{Pleural Effusion}.

\clearpage

\subsection{Performance across datasets based on Sex}
\begin{figure}[h]
    \centering
    \includegraphics[width=\linewidth]{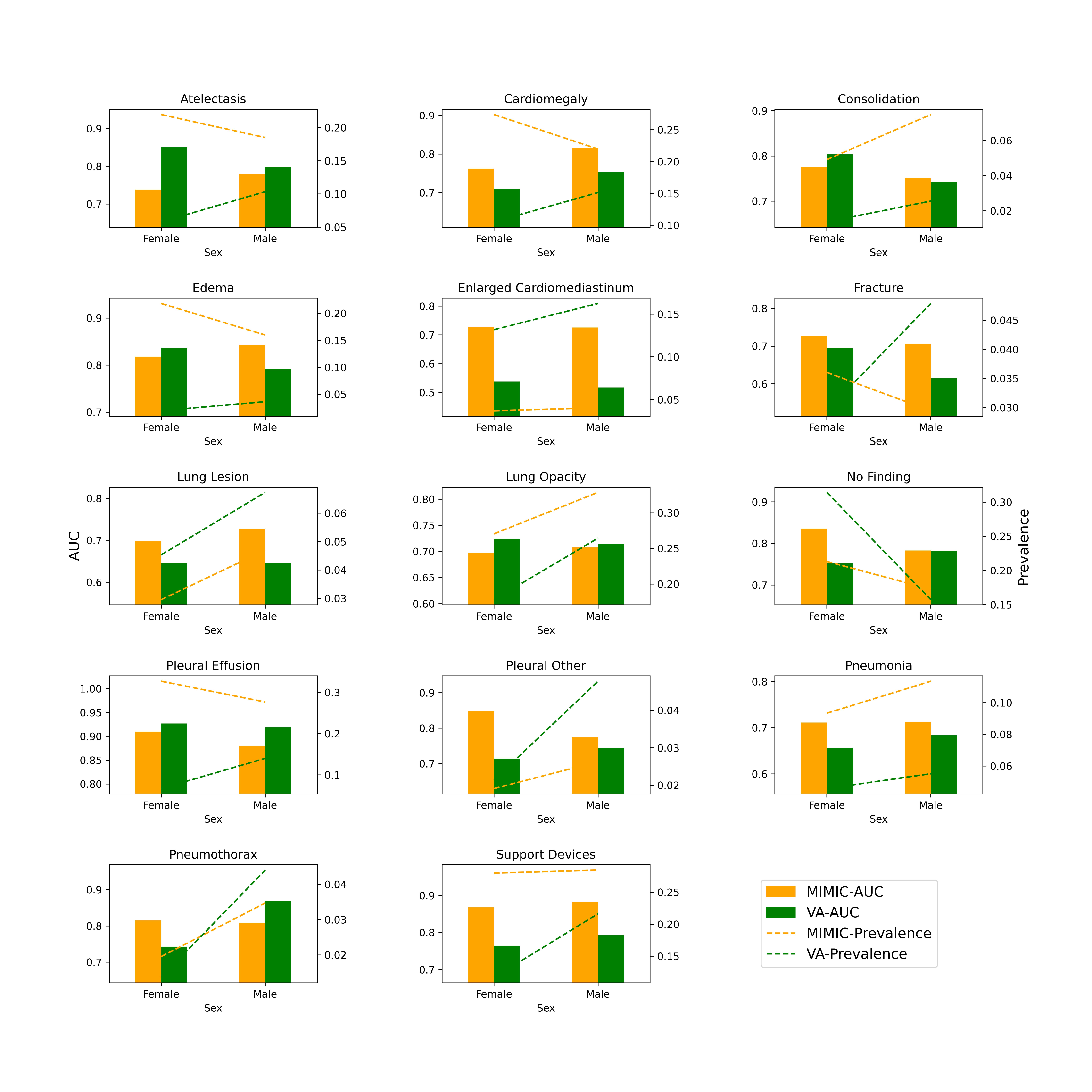}
     \caption{\centering Comparison of Label-wise AUC and Prevalence across the Sexes for MIMIC-CXR and VA-CXR. \footnotesize Orange represents MIMIC-CXR and Green represents VA-CXR}
     \label{fig:genderAcc}
\end{figure}

Figure~\ref{fig:genderAcc} compares the label-wise AUC and prevalence of the two sexes across the Test Split MIMIC-CXR and VA-CXR. This comparison is essential as the female-male patient ratio in VA-CXR is higher than that of MIMIC-CXR; dashed lines in the figure~\ref{fig:genderAcc} can observe this. 

\clearpage

\subsection{View Position-wise across datasets}
\begin{figure}[h]
    \centering
    \includegraphics[width=\linewidth]{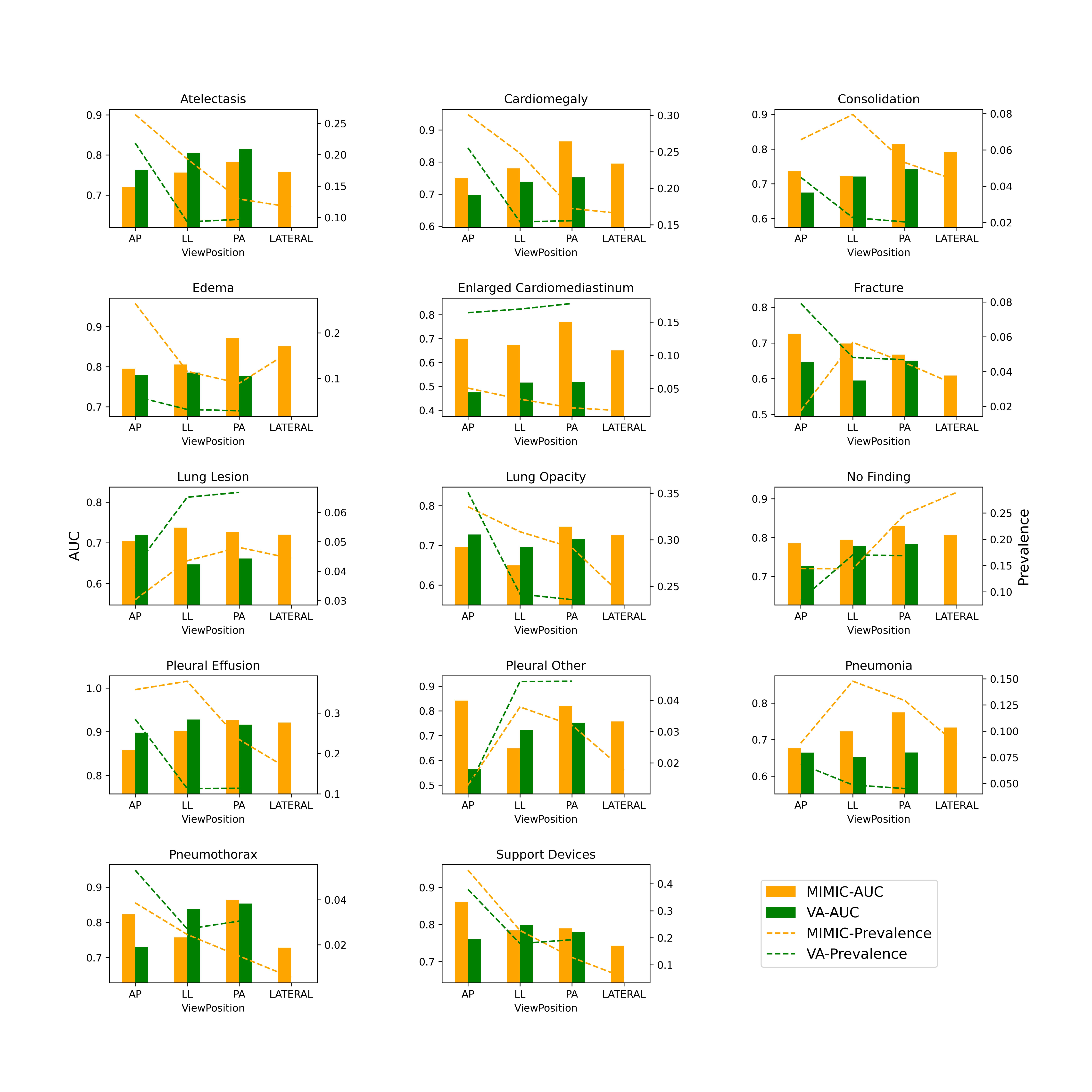}
    \caption{Comparison of AUC and Prevalence across View Points}
    \label{fig:viewPosition}
\end{figure}

Figure~\ref{fig:viewPosition} compares labels across Test-Split MIMIC-CXR and VA-CXR. The VA-CXR doesn't contain any \textit{Lateral} images; the figure shows only MIMIC-CXR performance on the \textit{Lateral}. The prevalence and AUC of viewpoints vary based on the label, with \textit{ECM} having the most difference between the datasets. \textit{Pleural Other} has a drop in performance in VA-CXR in \textit{AP} view position, potentially due to low prevalence in both datasets. 
 
\clearpage

\subsection{Age Group-wise comparison across datasets}
\begin{figure}[h]
    \centering
    \includegraphics[width=\linewidth]{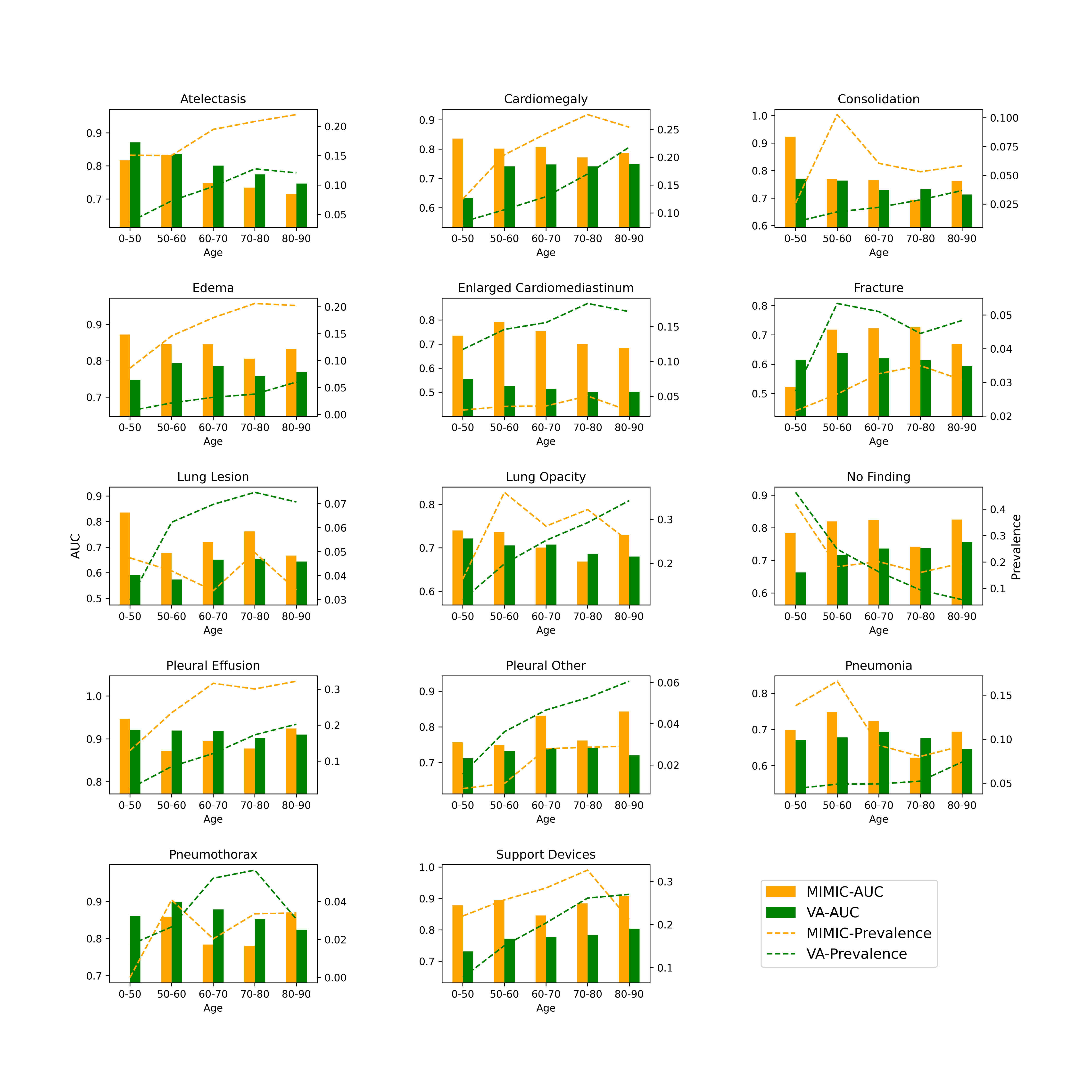}
    \caption{Age Group Label Wise Distribution}
    \label{fig:age-group}
\end{figure}

Figure~\ref{fig:age-group} shows the VA-CXR prevalence increase as the age increases across all labels except \textit{No Finding}, \textit{Lung Lesion}, and \textit{Pneumothorax}.  The highest performance drops of 0.15 to 0.2 AUC  in VA-CXR compared to the Test Split of MIMIC-CXR can be observed in \textit{Enlarged Cardiomediastinum} and \textit{Support Devices}. In VA-CXR, it is interesting that the AUC performance across the age groups is more stable, i.e., there is not much change in AUC, with the exception of \textit{Atelectasis}.  

\clearpage

\subsection{Summary}
As seen from the results, the ground truth validation and multi-label classification performance across the NLP extraction tools showed that though the VA-CXR dataset has lesser disagreement rates than the MIMIC-CXR datasets, and there were AUC differences between models when using ChexPert and ChexBert (potentially propagated by distribution differences). When comparing the multi-label classification performance on different datasets, the unseen datasets didn't show domain shift other than a few labels such as \textit{Enlarged Cardiomediastinum}. Among the different subgroup analyses, the study year had the most drastic differences in the performance of the multi-label classification model. These differences indicate that the domain shift is definitely of concern in study years.

\section{Discussion}

Our study quantified the domain shift between a publicly available critical care dataset (MIMIC-CXR) and an outpatient, private institutional dataset (VA-CXR), assessing the efficacy of transfer learning for chest X-ray classification.  Because domain shift encompasses multifaceted factors, our comprehensive study was structured into three interrelated parts: the quality of ground truth, performance comparison, and subgroup analysis. 

\textit{Quality of Ground Truth} 
Supervised learning models heavily rely on the quality of labels for training. Therefore, we quantified the ground truth quality extracted from radiology reports in both datasets. Despite being a source dataset, MIMIC-CXR exhibited significantly higher disagreement rates between the two ground truth information extraction NLP algorithms than the target, VA-CXR dataset. The potential performance drift of the target can be attributed to the mismatch of ground truth extraction methods. Our analysis underscored the importance of high-quality annotations in mitigating the effects of domain shift \cite{seyyed2020chexclusion,rajpurkar2023current} and improving model performance.

\textit{Performance Comparison:} 
We compared the difference in classification performance between MIMIC-CXR and VA-CXR to understand the extent of domain shift and its implications for model generalization. Our findings revealed notable variability in classification accuracy, highlighting the challenges posed by domain shift. We observe that the prevalence and performance were directly associated; for example, MIMIC-CXR's \textit{enlarged cardio mediastinum} low prevalence may be the reason for its low performance. 

\textit{Subgroup analysis} 
Subgroup analysis using demographic factors is crucial as we expect our populations to have different demographic distributions \cite{gichoya2022ai}. 
We observed a decline in the performance of the VA-CXR dataset over time, particularly in the years following 2020. This drop may be due to differences in the study years between the source dataset (MIMIC-CXR) and the VA-CXR dataset, as well as potential impacts from the pandemic years (note: we did not evaluate additional labels related to the pandemic).
We observed that both groups, though with different distributions, have aging populations across conditions of interest. Still, the prevalence of most conditions with age increases in the VA-CXR dataset, but it is highly variable in MIMIC-CXR.  For sex subgroup analysis, the performance was similar across the female and male populations in the datasets. Though the VA-CXR population is skewed towards the male population, the model performed well on the female population. This can be attributed to the source domain (MIMIC-CXR) having a balanced male-to-female ratio. 
For the view-based subgroup, we observed high variability between the views. This behavior can be attributed to the necessity of specific views for accurately diagnosing certain diseases. For instance, conditions such as Pneumonia and ECM primarily rely on PA (Posteroanterior) or AP (Anteroposterior) views for diagnosis.

Chest radiography is a first-line imaging modality for assessing lung and heart conditions. With advanced artificial intelligence techniques, the automation of chest X-ray abnormality detection has seen significant progress \cite{irmici2023chest}. This has accelerated the pace of diagnosis and treatment and opened avenues for exploring biomedical applications in various clinical settings. However, the efficacy of these machine learning models heavily relies on the quality and availability of training data \cite{soin2022chexstray}.

\textit{Clinical relevance}
Despite the surge in artificial intelligence and machine learning applications in clinical settings, their adoption has been uneven, primarily due to disparities in funding\cite{khan2023drawbacks}, expertise, and availability of computing resources. While academic medical centers have been at the forefront of AI adoption, the accessibility of such resources remains a challenge for smaller research institutions. Introducing transfer learning has mitigated some of these challenges by enabling researchers to leverage pre-existing models trained on publicly available datasets for their specific applications. However, transfer learning introduces the concept of domain shift, which poses a significant obstacle to achieving robust and generalizable models.

\textit{Potential Workflows}
While we primarily focused on matching image classification to known findings typical of chest X-rays and extracted those from radiology reports using NLP, this may not be the best workflow with which image classification can be incorporated into a medical outcomes workflow.  Tiu, \textit{et al.} have noted the difficulties and inaccuracies associated with training chest X-ray classifiers against such labels and demonstrated improved accuracy and efficiencies with a self-supervised workflow to train their image classification model.\cite{tiu22}.  Even more ambitiously, they argue in Reference \cite{moor23} that all modalities of data should be co-optimized in a much broader framework, known as a foundation model.  Our observations in Figure \ref{fig:sensDx.png} that most of the X-ray findings do not map directly onto diagnoses support the idea that more flexible workflows may be required to optimally incorporate AI-based image classification into medical outcome prediction.

By systematically addressing the challenges of domain shift and demographic factors, our study contributes valuable insights to developing more accurate and robust chest X-ray classification models. Furthermore, our findings have broader implications for healthcare, emphasizing the importance of understanding and mitigating biases in machine learning applications to ensure equitable healthcare delivery. Our research serves as a foundation for future studies aimed at refining transfer learning techniques and enhancing the generalizability of machine learning models in clinical practice.

\textit{Limitations}
It is crucial to acknowledge that the data for the two studies were collected under significantly different clinical settings, which may influence the outcomes. The MIMIC database primarily includes inpatient data, where patients may be more critically ill, as indicated by the prevalence of portable anteroposterior (AP) images that tend to be less precise. In contrast, the VA outpatient population likely represents individuals who are relatively less ill, with higher-quality image data collection due to the setting's focus on outpatient care. Although we were unable to assess these differences quantitatively, we hypothesize that the clinical setting could substantially impact the performance of our models, influenced by the severity of diseases and other contextual factors.

\section{Conclusion}
In conclusion, our study sheds light on the critical interplay between domain shift, demographic factors, and the efficacy of transfer learning in chest X-ray classification. By quantifying the domain shift between a publicly available dataset and a private institutional dataset, we have identified disparities in classification accuracy and highlighted the challenges posed by transferring models across domains. Moreover, our analysis of demographic factors underscores the importance of considering population diversity in model development to ensure equitable healthcare outcomes. Moving forward, addressing these challenges will require concerted efforts to improve data quality, develop robust transfer learning techniques, and enhance the generalizability of machine learning models in clinical practice. Ultimately, by fostering a deeper understanding of these complexities, our research paves the way for the development of more accurate, reliable, and accessible chest X-ray classification models with the potential to transform patient care and advance the field of medical imaging.

\section{Acknowledgments}
This work is sponsored by the US Department of Veterans Affairs using resources from the Knowledge Discovery Infrastructure which is located at the Oak Ridge National Laboratory and supported by the Office of Science of the U.S. Department of Energy. This manuscript has been authored by UT-Battelle, LLC, under contract DE-AC05-00OR22725 with the US Department of Energy (DOE). The US government retains and the publisher, by accepting the article for publication, acknowledges that the US government retains a nonexclusive, paid-up, irrevocable, worldwide license to publish or reproduce the published form of this manuscript or allow others to do so for US government purposes. DOE will provide public access to these results of federally sponsored research in accordance with the DOE Public Access Plan (
http://energy.gov/downloads/doe-public-access-plan
). 

\printbibliography

\clearpage

\onecolumn
\setcounter{figure}{0} \renewcommand{\thefigure}{A.\arabic{figure}}
\setcounter{table}{0} \renewcommand{\thetable}{A.\arabic{table}}

\section{Appendix:}

\subsection{Age When Image was Taken versus Study Year}

\begin{figure}[h]
    \centering
    \includegraphics[width=\linewidth]{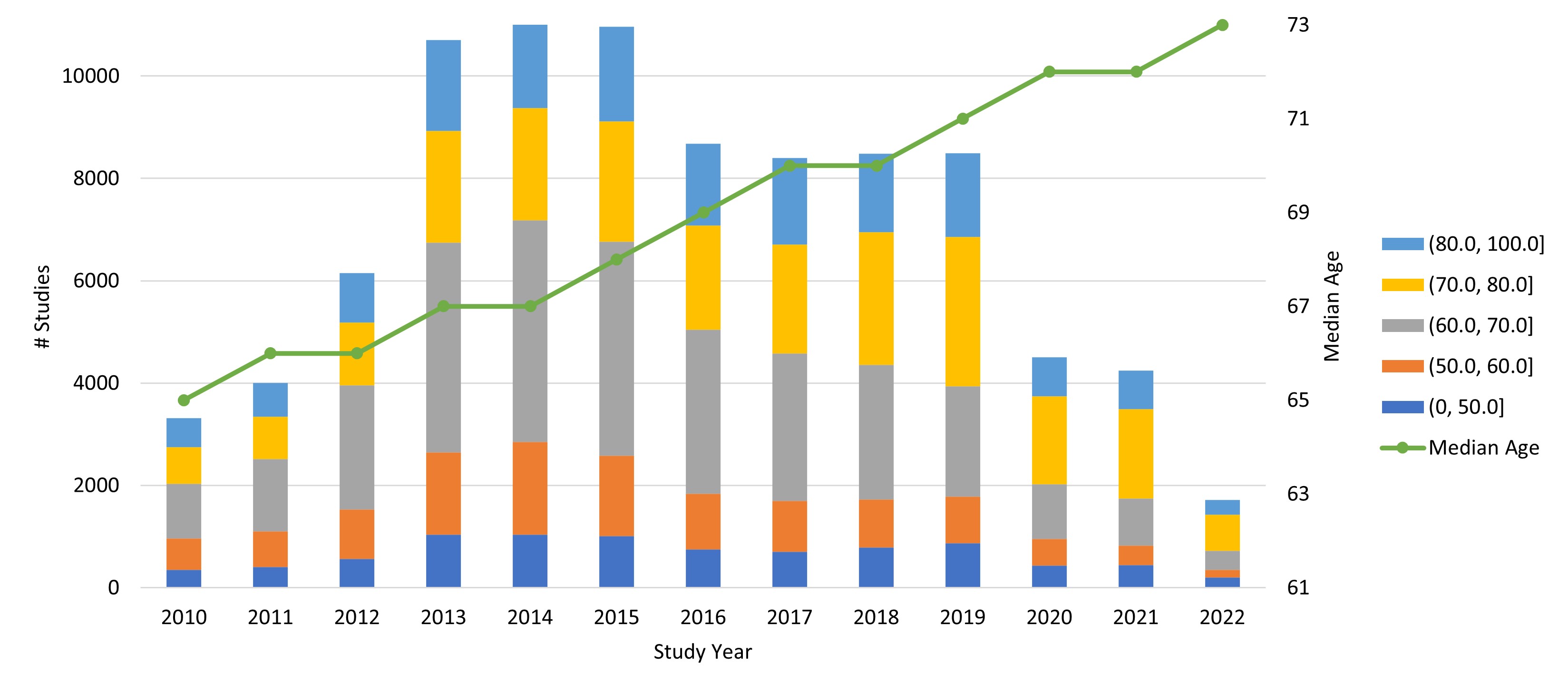}
    \caption{Study Year vs AgeWhenImageWasTaken}
    \label{fig:ageWhenImage}
\end{figure}

\begin{table}[h]
    \centering
    \begin{tabular}{|c||cc||cc|}
    \hline
    & \multicolumn{2}{|c|}{\textbf{Source Dataset}}      & \multicolumn{2}{|c|}{\textbf{Target Dataset}}  \\ \hline
    \textbf{Name}         & \multicolumn{2}{|c|}{MIMIC-CXR}              & \multicolumn{2}{|c|}{VA-CXR} \\ \hline
       \textbf{View Position} &  \# Images & \% & \# Images & \% \\ \hline
        Anterior/Posterior (AP) & 147,169 & 39.02 & 879 & 0.34\\ \hline
        Posterior/Anterior (PA) & 96,155 & 25.5 & 52,749 & 20.33 \\ \hline
        Lateral & 82,852 & 21.97 & 11 & - \\ \hline
        Left Lateral (LL) & 35,129 & 9.31 & 50,267 & 19.38\\ \hline
        Others & 23 & - & 6 & - \\ \hline
        Unknown & 15,769 & 4.18 & 159,482 & 61.49 \\ \hline
    \end{tabular}
    \caption{View Position Distribution between datasets}
    \label{Apptab:viewPosition}
\end{table}

\begin{table}[]
    \centering
    \begin{tabular}{|c|c|c|} \hline 
         \textbf{Finding}&  \textbf{ICD-9 Codes}&\textbf{ICD-10 Codes}\\ \hline 
         Pneumonia& 
    480 to 486, 487.0&J10.0, J11.0, J2 to J8\\ \hline 
 Atelectasis&518.0, 770.4&J98.11\\ \hline 
 Cardiomegaly&429.3&I51.7\\ \hline 
 Edema& 518.4, 514, 428&J81, I50\\ \hline 
 Enlarged Cardiomediastinum& 519.2&J98.5\\ \hline 
 Fracture& 807.[0to4], 806.[2to3]&S22\\ \hline 
 Lung Lesion& 793.11&R91\\ \hline 
 Lung Opacity& 516&J84\\ \hline 
 Pleural Effusion& 511.[1,8,9]&J90,J91\\ \hline 
 Pleural Other& 511.0, 515&J92,J94,J84.[0,1]\\ \hline 
 Pneumo thorax& 512.[0,1,8], 512.8[1,2,3,9]&J93\\ \hline\end{tabular}
    \caption{Groundtruth Validation using ICD Codes}
    \label{tab:ground_icd}
\end{table}

\begin{table}
\centering

\begin{tabular}{|p{2cm}|p{3cm}| l| l| l| l| l| l| l| l|}        \hline
 \multirow{2}{*}{Findings}&  \multirow{2}{=}{ICD Dx w.r.t Img}& \multicolumn{2}{l|}{Uncertain} & \multicolumn{2}{l|}{Negative} & \multicolumn{2}{l|}{Positive} & \multicolumn{2}{l|}{Null} \\ \cline{3-10}
 &  & Bert & Pert & Bert & Pert & Bert & Pert & Bert & Pert \\ \hline
\multirow{3}{=}{Pneumonia} & ICD Dx \textless1 wk & 841 & 762 & 248 & 91 & 1336 & 1590 & 2808 & 2790 \\ \cline{2-10}
 & ICD Dx  \textgreater 1 wk & 1895 & 1546 & 1366 & 675 & 1715 & 2821 & 29002 & 28936 \\ \cline{2-10}
 & No Dx & 1261 & 1018 & 1657 & 965 & 564 & 1571 & 51964 & 51892 \\ \hline
\multirow{3}{=}{Atelectasis} & ICD Dx  \textless 1wk & 146 & 138 & 3 & 2 & 415 & 424 & 363 & 363 \\ \cline{2-10}
 & ICD Dx  \textgreater 1 wk & 692 & 680 & 9 & 15 & 1366 & 1375 & 5068 & 5065 \\ \cline{2-10}
 & No Dx & 3983 & 3879 & 45 & 83 & 7573 & 7626 & 71794 & 71807 \\ \hline
\multirow{3}{=}{Cardio- megaly} & ICD Dx  \textless 1wk & 59 & 51 & 82 & 54 & 469 & 477 & 187 & 215 \\ \cline{2-10}
 & ICD Dx  \textgreater 1 wk & 887 & 821 & 2098 & 1499 & 1889 & 2220 & 3651 & 3985 \\ \cline{2-10}
 & No Dx & 6121 & 5745 & 31441 & 22643 & 6925 & 10967 & 37627 & 42759 \\ \hline
\multirow{3}{=}{Edema} & ICD Dx 90 days & 2054 & 2154 & 6859 & 6247 & 1677 & 2057 & 8271 & 8403 \\ \cline{2-10}
 & ICD Dx \textgreater 90 days & 2456 & 2629 & 13590 & 12653 & 1806 & 2312 & 16886 & 17144 \\ \cline{2-10}
 & No Dx & 833 & 973 & 20304 & 19353 & 403 & 702 & 32598 & 33110 \\ \hline
\multirow{3}{=}{ECM} & ICD Dx  \textless 1wk & 32 & 23 & 5 & 3 & 7 & 17 & 54 & 55 \\ \cline{2-10}
 & ICD Dx  \textgreater 1 wk & 136 & 78 & 48 & 44 & 29 & 96 & 266 & 261 \\ \cline{2-10}
 & No Dx & 23755 & 14535 & 15744 & 3992 & 3091 & 14576 & 47924 & 47411 \\ \hline
\multirow{3}{=}{Fracture} & ICD Dx  \textless 1wk & 0 & 1 & 2 & 0.5 & 18 & 18 & 297 & 298 \\ \cline{2-10}
 & ICD Dx  \textgreater 1 wk & 1 & 1 & 38 & 24 & 238 & 244 & 4278 & 4286 \\ \cline{2-10}
 & No Dx & 96 & 152 & 962 & 652 & 3913 & 4029 & 81262 & 81401 \\ \hline 
\multirow{3}{=}{Lung Lesion} & ICD Dx  \textless 1wk & 5 & 5 & 29 & 14 & 162 & 168 & 4251 & 4260 \\ \cline{2-10}
 & ICD Dx  \textgreater 1 wk & 37 & 52 & 308 & 185 & 1502 & 1546 & 30315 & 30379 \\ \cline{2-10}
 & No Dx & 60 & 101 & 692 & 490 & 2632 & 2708 & 54768 & 54853 \\ \hline
\multirow{3}{=}{Lung Opacity} & ICD Dx  \textless 1wk & 0 & 31 & 60 & 46 & 301 & 424 & 342 & 202 \\ \cline{2-10}
 & ICD Dx  \textgreater 1 wk & 10 & 246 & 606 & 761 & 1852 & 2444 & 3131 & 2148 \\ \cline{2-10}
 & No Dx & 178 & 2748 & 11179 & 18518 & 16136 & 21049 & 57794 & 42972 \\ \hline
\multirow{3}{=}{Pleural Effusion} & ICD Dx  \textless 1wk & 201 & 277 & 326 & 276 & 3127 & 2970 & 619 & 750 \\ \cline{2-10}
 & ICD Dx  \textgreater 1 wk & 518 & 1098 & 5580 & 5080 & 5343 & 5060 & 2971 & 3174 \\ \cline{2-10}
 & No Dx & 1238 & 4015 & 55570 & 52665 & 6719 & 6665 & 12141 & 12323 \\ \hline
\multirow{3}{=}{Pleural Other} & ICD Dx  \textless 1wk & 83 & 71 & 0.5 & 2 & 352 & 307 & 3838 & 3893 \\ \cline{2-10}
 & ICD Dx  \textgreater 1 wk & 212 & 216 & 1 & 9 & 1270 & 1089 & 12929 & 13098 \\ \cline{2-10}
 & No Dx & 422 & 462 & 11 & 31 & 3710 & 3109 & 71525 & 72066 \\ \hline
\multirow{3}{=}{Pneumo thorax} & ICD Dx  \textless 1wk & 57 & 42 & 303 & 265 & 782 & 834 & 139 & 140 \\ \cline{2-10}
 & ICD Dx  \textgreater 1 wk & 59 & 114 & 1168 & 1037 & 775 & 836 & 1556 & 1571 \\ \cline{2-10}
 & No Dx & 250 & 1871 & 35272 & 32893 & 2317 & 2739 & 49095 & 49431 \\ \hline

\end{tabular}
 \caption{Groundtruth Validation using ICD Codes Count \footnotesize{ECM- Enlarged Cardiomediastinum; Dx - Diagnosis;  ICD Dx w.r.t Img - ICD Diagnose Code with respect to the Date of imaging study  }}
    \label{tab:ground_count}
\end{table}

\end{document}